\documentclass{aa}  

\usepackage{graphicx}
\usepackage{txfonts}
\usepackage[english]{babel}
\usepackage{natbib}
\begin{document}

   \title{Photometric survey of 67 near-Earth objects}


     \author{ S. Ieva \inst{1},
          E. Dotto \inst{1},
          E. Mazzotta Epifani \inst{1},
          D. Perna \inst{1,2},
          A. Rossi \inst{3}, 
           M. A. Barucci \inst{2},
         A. Di Paola \inst{1},
          R. Speziali \inst{1},
         M. Micheli \inst{1,4},
         E. Perozzi  \inst{5},
         M. Lazzarin \inst{6},
          I. Bertini \inst{6}            
}


\institute{ INAF -  Osservatorio Astronomico di Roma, Via Frascati 33, 00078 Monte Porzio Catone, Rome, Italy\\
        \email{simone.ieva@oa-roma.inaf.it}
\and LESIA - Observatoire de Paris, PSL Research University, CNRS, Sorbonne Universit\'es, UPMC Univ. Paris 06, Univ. Paris Diderot, Sorbonne Paris Cit\'e, 5 place Jules Janssen, F-92195 Meudon, France
\and IFAC- CNR, Via Madonna del Piano 10, 50019 Sesto Fiorentino, Firenze, Italy
\and ESA SSA-NEO Coordination Centre, Largo Galileo Galilei, 1, 00044, Frascati (RM), Italy
\and Agenzia Spaziale Italiana, Via del Politecnico 1, 00100, Rome, Italy
\and Department of Physics and Astronomy `Galileo Galilei', University of Padova, Vicolo dell'Osservatorio 3, I-35122 Padova, Italy 
}

   \date{Received 23 October 2017 / Accepted 29 March 2018. }

 
  \abstract
   {The near-Earth object (NEO) population is a window into the original conditions of the protosolar nebula, and has the potential to provide a  key pathway for the delivery of water and organics to the early Earth. In addition to  delivering the crucial ingredients for life, NEOs can pose a serious hazard to humanity since they can impact the Earth. To properly quantify the impact risk, physical properties of the NEO population need to be studied. Unfortunately, NEOs have a great variation in terms of mitigation-relevant quantities (size, albedo, composition, etc.) and less than 15\% of them have been characterized to date.}
   {There is an urgent need to undertake a comprehensive characterization of smaller NEOs (D<300m) given that there are many more of them than larger objects; their small sizes make them intrinsically fainter and therefore harder to study. One of the main aims of the NEOShield-2 project (2015--2017), financed by the European Community in the framework of the Horizon 2020 program, is therefore to retrieve physical properties of a wide number of NEOs in order to design impact mitigation missions and assess the consequences of an impact on Earth. }
   { We carried out visible photometry of NEOs, making use of the DOLORES instrument at the Telescopio Nazionale Galileo (TNG, La Palma, Spain) in order to derive visible color indexes and the taxonomic classification for each target in our sample.  }
%
   {We attributed for the first time the taxonomical complex of 67 objects obtained during the first year of the project.  While the majority of our sample belong to the S-complex, carbonaceous C-complex NEOs deserve particular attention.  These NEOs can be located in orbits that are challenging from a mitigation point of view, with high inclination and low minimum orbit intersection distance (MOID).  In addition, the lack of  carbonaceous material we see in the small NEO population might not  be due to an observational bias alone. 
}
   {}

   \keywords{Minor planets, asteroids: NEOs, Techniques: photometric, Surveys}
   
\titlerunning{A photometric survey of 67 NEOs}
\authorrunning{S. Ieva et al.}

   \maketitle
%

\section{Introduction}

The near-Earth object (NEO) population has become very important in the last thirty years;  it has been acknowledged to represent the most accessible vestiges of the building blocks that formed the solar system approximately some 4.5 billion years ago. The study of NEOs furthers the understanding of the initial conditions in the protosolar nebula and sets important constraints on the formation of the  solar system in an era when exoplanet discoveries seem to have complicated the classical scenario of planetary formation (Winn \& Fabrycky, \citeyear{winn2015}). Furthermore, NEOs can help us  answer fundamental questions about the presence  of water and organics on the early Earth, and  last but not least, of life itself. Recent astrobiological studies suggest that it is plausible that comets and NEOs are responsible for the delivery of organic and prebiotic molecules to the Earth (Izidoro et al. \citeyear{izidoro2013}). 

The study of the  physical characteristics of NEOs is also compelling in view of the potential hazard posed to our planet. NEOs are linked with all kinds of meteorite falls, from the recent Chelyabinsk event (Popova et al. \citeyear{popova2013}) to the occasional catastrophic impact events (like the K-T event, Rehan et al. \citeyear{rehan2013}). In case of possible impactors their physical characterization is crucial to defining successful mitigation strategies  (see, e.g., Perna et al. \citeyear{perna2013}). Unfortunately, more than 85\% of the $\sim$18.000 known NEOs\footnote{https://newton.dm.unipi.it/neodys}   still lack a compositional characterization, and their increasing discovery rate (currently 1.900 objects/year) makes the situation progressively worse.

A broad range of diversity in terms of composition and spectral properties is also present among the NEO population. All the taxonomic classes in the main belt, the predominant source region for NEOs, are represented in the distribution of NEOs taxonomic types. However, the S-complex is by far the most common type of NEOs observed, while at
the moment the C-complex objects only account   for  15\% of the taxonomic distribution (Binzel et al. \citeyear{binzel2015}). The underrepresentation of the C-complex
 and generally of low albedo NEOs is even more unusual considering that they represent the majority of the main belt population. There is a growing evidence that this could be due to an observational bias among taxonomic types that favors the discovery of small and bright silicate asteroids rather than big and dark carbonaceous asteroids. Thermal infrared surveys, like NEOWISE, suggest  that ~35\% of all NEOs discovered have low albedos (Mainzer et al. \citeyear{mainzer2015}). 
Furthermore, \citet{delbo2014} has shown that even thermal fragmentation can destroy preferentially dark bodies, reducing faint and dark carbonaceous material to a size limit below which they are no longer detected.
 
NEOs below a few hundred meters deserve attention since they greatly outnumber the larger objects, thus increasing their chances of impact with Earth.   Even an object that is a few hundred meters in diameter is capable of causing severe regional damage (Perna et al. \citeyear{perna2015}). Because of the potential threat to human civilization posed by NEOs, several space agencies and international organizations
are currently studying how to plan in advance possible countermeasures that could mitigate an NEO impact.
At the European level, the European Commission has promoted the study of NEOs by approving and financing   the NEOShield-2 project (2015--2017) in the framework of the Horizon 2020 program. 
One of its main aims is to undertake for the first time ever a comprehensive characterization of the small NEO population, and to study their typical mitigation-relevant physical properties (size, albedo, mineralogy, shape, density, internal structure) in the occurrence of a future mitigation mission toward one of these objects.
In particular, the Italian team has been in charge of the characterization of the NEO population via visible photometry.

In this article we present photometric color indexes and a taxonomic classification for a sample of 67 objects observed at the Telescopio Nazionale Galileo (TNG). In the following sections we report the results we obtained during the first year of the NEOShield-2 project: in section 2 we describe the observational settings and data reduction procedures; in section 3 we present the results we obtained in our sample using different complementary techniques; in section 4 we discuss our findings in the light of recent advancement in the field; and finally in section 5 we discuss our conclusions.

\section{Observations and data reduction}
The data presented in this work were collected between September 2015 and September 2016 at the Telescopio Nazionale Galileo (TNG, La Palma, Canary Islands, Spain).  Targets were selected from  the NEOs observable each night using a prioritization algorithm (Cortese et al. \citeyear{cortese2017}), optimized for the NEOShield-2 requirements. At present, they all have no physical characterization according to the EARN\footnote{www.earn.dlr.de/nea/table1\textunderscore{new}.html}   database. 
In order to increase our sample we included in our analysis previous observations for seven objects obtained at TNG in 2014. Observational conditions are given  in Table 1.

Visible photometry was performed with the Device Optimized for the LOw RESolution (DOLORES) instrument, equipped with a 2048 x 2048 E2V 4240 thinned back-illuminated, deep-depleted, Astro-BB coated CCD with a pixel size of 13.5 $\mu$m. We used the broadband B-V-R-I filters, adjusting the exposure time according to the object magnitude in order to obtain a S/N > 40, thus reaching a level of precision in the determination of the magnitude able to differentiate taxonomy in visible wavelengths. 

Typically, we used a B-V-R-V-I photometric sequence, repeating the V filter twice to avoid systematic errors.
Images were reduced with the MIDAS software package using standard techniques  (see, e.g., Perna  et al. \citeyear{perna2010}): bias subtraction and flat field correction;
measurement of instrumental magnitude via aperture photometry by integrating on a radius  about three times the average seeing, and removing the sky contribution using a 5--10 pixel annulus around each object.
Absolute calibration was performed by observing several standard stars each night (Landolt et al. \citeyear{landolt1992}). The error bars were computed taking into account the photometric errors and the instrumental magnitude statistics. 

We properly reduced and calibrated B-V-R-I photometric measurements for a total sample of 67 individual NEOs. At the moment of writing the present article, there are  no colors or taxonomy reported in the literature for any of them. The obtained color indexes are listed in Table 2. Seven NEOs were observable during two slots and were observed twice.  We checked them for possible color variations caused by rotational light curves.  Since the obtained color indexes do not present any intrinsic variation (see Table 2), we considered average values in the subsequent analysis.

\section{Data analysis}
\subsection{Taxonomy}
Taxonomic classification based upon photometric observations is a powerful tool for investigating the surface composition of a large sample of NEOs with a limited amount of observing time. 

Starting from the photometric B, V, R, and I fluxes, first we obtained the B-V, V-R, and V-I color indexes given in Table 2. Then, in order to compare our color indexes with reflectance sample spectra  from \citet{demeo2009}, we normalized to unity the reflectance at the V filter, and we rescaled the B, R, and I reflectance using \\

$\mathcal{R}_{\lambda} = 10^{-0.4[(M_{\lambda}-V)-(M_{\lambda}-V)_\odot]}$,  \\

\noindent
where $(M_{\lambda}-V)$ and $(M_{\lambda}-V)_\odot$ are the colors for the object and the Sun at the wavelength $\lambda$.

We taxonomically classified our sample using the M4AST online classification algorithm (Popescu et al. \citeyear{popescu2012}) that adopts standard curve matching techniques to determine to what extent the reflectance of an asteroid is similar to a standard spectrum, considering each class portrayed in \citet{demeo2009}.
For our analysis we considered three major groups: the S-complex (including S-, Sa-, Sq-, Sr-, Sv-, and Q-type objects), the C-complex (B-, C-, Cb-, Cg-, Cgh-, and Ch-type objects), and the X-complex (X-, Xc-, Xe-, Xk-, and Xn-type objects). We also classified a few NEOs as belonging to several end-members (A-,  D-,  and V-type objects). The final taxonomic classification is listed in Table 2.

In order to have an independent confirmation of the taxonomy, we compared the  B-R (B-V + V-R) and V-I color indexes that we obtained for NEOs in our sample with the same colors derived using the sample spectra for each class retrieved by \citet{demeo2009}. The latter were obtained integrating the flux of these average spectra over the transmission range of each B-V-R-I filter.

In Fig. 1 we show the B-R versus V-I color-color diagram for NEOs observed in our survey. Values from the sample spectra of \citet{demeo2009} are shown in bold letters. It is clear that different taxonomic groups occupy different regions of the diagram and that our taxonomic classification is reliable. A general agreement for C, S, and X-complex, and for D-types is evident.  The great intrinsic spectral variation found in the visible range for basaltic material (see, e.g., Ieva et al. \citeyear{ieva2016}) could be responsible for the great dispersion for V-type objects.
Due to the presence of only one NEO belonging to the A-type in our survey, no reliable comparison is possible at the moment.

We also present in Table 3  the B-R and V-I median colors and the 1-$\sigma$ deviation for the different complexes in our sample, together with the values obtained from the DeMeo spectra. Error bars in this case take into account the dispersion of each sample spectrum. 

\subsection{Taxa versus orbital parameters}

To investigate the relationship between taxonomic classification (and therefore surface composition) and dynamical properties, we analyzed the distribution of different taxa in our sample according to their orbital parameters: semimajor axis \emph{a}, eccentricity \emph{e}, inclination \emph{i}, perihelion \emph{q}, aphelion \emph{Q}, and minimum orbit intersection distance ({MOID}) with our planet. The last in particular allows us to define the  potentially hazardous asteroids (PHAs), which are objects with a present MOID with the Earth < 0.05 au and an absolute magnitude H < 22. PHAs are therefore NEOs passing close to our planet that are big enough to cause regional or even global damage. Assuming the average albedo found by Mainzer et al. \citeyear{mainzer2011b}, these objects could be bigger than 140 m, although recently it has been proposed to include objects large enough to penetrate the Earth's atmosphere\footnote{cneos.jpl.nasa.gov/doc/SDT\textunderscore{report}\textunderscore{2017}.html}. 
Therefore, the characterization of the composition and the mitigation-relevant  quantities for objects down to 30 m is crucial. With our survey, we performed for the first time the taxonomical classification for 23 PHAs. They are listed  with an asterisk in Table 2.

Median values of the orbital parameters are shown in Table 4 for the two most represented taxonomic classes in our sample: the C- and S-complexes.  Silicate S-complex objects have the lowest median values for both \emph{a} and  \emph{q}, as expected for objects formed in the inner part of the main belt. They also show the lowest average {MOID} in the sample. Carbonaceous asteroids exhibit higher \emph{a} and  \emph{Q} values, as   expected for objects originated from the middle and outer parts of the main belt (Morbidelli et al. \citeyear{morbidelli2002}).

In Fig. 2 we show the orbital inclination versus Earth's  MOID  for all the objects characterized in our sample, with a focus on the PHA population. The majority of PHAs in our survey (15) belongs to the S-complex, in agreement with the recent results of \citet{perna2016}. 

Objects with a low MOID can have a close encounter with our planet, and the lower the MOID, the higher is the impact risk in the next centuries. Moreover, this parameter can be altered  by close encounters with terrestrial planets, and can change up to $\pm $ 0.05 au over a century. For this reason, objects with an extremely low MOID should be monitored in the next decades.
In addition, the accessibility of a NEO from Earth  can be measured by the $\Delta V$, the velocity increment required for a spacecraft to reach the asteroid orbit. This parameter can be used as a reliable estimation of the fuel necessary to arrive at the target. 
Current propulsion technologies limit the accessible NEOs to those with  $\Delta V$ below 7 km/s (Hinkle et al. \citeyear{hinkle2014}). Since NEOs on orbits with an inclination \emph{i}  > 10$^{\circ}$ have greater $\Delta V$, they are at the moment inaccessible from our planet.

Two C-complex bodies in our survey (154275 and 430804)   show very low MOID (< 0.04 au) and very high inclination ( > 10$^{\circ}$).
Moreover, at present the most advanced mitigation technique, the kinetic impactor, strongly depends on the porosity of the impacting objects; it seems  more challenging to  execute this technique on a C-type rather than a S-type asteroid (Perna et al. \citeyear{perna2013}, Drube et al. \citeyear{drube2015}). Therefore, the more porous C-complex objects in low MOID, high inclination orbits represent at the moment a greater risk in terms of mitigation, and should require in the future a more detailed investigation (e.g., near-IR spectroscopy). 

\subsection{Statistical analysis}
In our survey we classified for the first time 67 uncharacterized NEOs. We found that the  majority belong to the S-complex (41), while C- and X-complex account for 11 and 7 objects, respectively. We also identified four V-type objects, three D-type bodies, and only one A-type asteroid. The taxonomic distribution of the observed NEOs in our sample is comparable with the distribution of taxonomic types retrieved from the EARN database (See Fig. 3a). However, since we noticed an  overrepresentation of C-complex bodies (and an underrepresentation of silicate S-complex targets) with respect to the EARN database, we decided to analyzed the distribution of these taxonomic classes in greater  detail.

The absolute magnitude H can be used as a proxy to determine the target size once the geometrical albedo is known. The knowledge of asteroid diameters is crucial in order to constrain the impact energy of NEOs and therefore to plan a proper mitigation strategy. Unfortunately, albedo has been computed only for a handful of NEOs (See Table 2 with the known albedos for targets in our sample). Therefore, to retrieve the size of the observed NEOs, whenever the real albedo was not available, we used the mean albedo of the derived taxonomic class (as given by Ryan \& Woodward \citeyear{ryan2010}).

A statistical bias could arise comparing C- and S-complex at a fixed H, since the lower albedo for carbonaceous objects results in a greater diameter. This seems particularly important for scarcely populated binnings of  H  (i.e., the extremes of the distribution).
In our present collection of NEOs we only have  four objects with H < 17 and one with H > 22. To increase the significance of our results, we decided to not consider these objects in the following statistical analysis.
We divided our remaining sample of 62 NEOs in three major groupings according to their estimated size: D < 300 m, 300 m < D < 800 m, and D > 800 m. These three ranges were chosen in order to have a comparable number of objects in each bin (22, 24, and 16, respectively). Despite the limited number of targets, our analysis still pointed out some interesting results that are discussed in the next section in a broader context. For example, the only A-type object found in our survey is in the small diameter range.
In addition, as shown in Fig. 3b,  S-complex bodies have a comparable number of objects in the small and medium range, with a slight drop in the large range; on the contrary, C-complex bodies  only have  one object in the small range, but a comparable number in the medium and large range.

\section{Discussion}

With our survey we increased by 9\% the number of physically characterized PHAs already known (See Perna et al. \citeyear{perna2016}). 
The physical characterization of this subclass of NEOs is particularly important since it can pose a serious hazard to human civilization. We confirmed that a large part of this population belongs to the S-complex group. 
However, the most dangerous objects in our sample are represented by porous C-complex targets with a low MOID and on high inclination (>10 deg) orbits. The current most advanced technique to prevent an impact hazard is the kinetic impactor (Drube et al. \citeyear{drube2015}), which strongly depends on the porosity of the threatening object. Moreover, to organize a mitigation mission towards objects coming from these orbits could be particularly challenging since the $\Delta V$ for these objects is greater than current technological capabilities. In addition, C-complex bodies usually have a lower albedo, which could result in a greater diameter than expected using an average albedo. So, for a proper evaluation of the mitigation-relevant quantities (e.g., size) these objects require a further analysis.

The taxonomic distribution in our sample is in agreement with literature (Binzel et al. \citeyear{binzel2015}) and the EARN database, with the overall majority of NEOs belonging to the S-complex, and only a few targets to the C-complex.
This could be explained by the efficiency of the transport mechanisms from the main belt towards the inner regions of the solar system, since S-complex are more common in the inner main belt and closer to highly efficient resonances (Binzel et al. \citeyear{binzel2004}), while C-complex are connected with the middle/outer main belt, as confirmed by our orbital analysis.

Our size analysis pointed out that there is a lack of carbonaceous asteroids in the small size range (D < 300 m).
Carbonaceous material generally has a low albedo, resulting in an observational bias that favors the discovery of larger/brighter carbonaceous asteroids rather than the smaller/fainter ones.  Although focused mostly on main belt asteroids, the analysis of
\citet{mainzer2012} has pointed out that, when comparing sizes and albedos for asteroids with taxononomic classification discovered by optical surveys, there seems to be a paucity of small, low albedo asteroids. However, when considering a IR-selected sample based on 12-micron flux, \citet{mainzer2011b} has found no significant variation in the ratio of bright to dark albedo asteroids from $\sim$ 10 km down to a few hundred meters. This could suggest that that observational selection effects can be at play when observing the lack of small, dark NEOs found by visible surveys. 

The apparent minority of C-complex objects with  decreasing size could also be related to the recent findings described in \citet{scheeres2017}. It is well known that the Yarkovsky - O'Keefe - Radzievskii - Paddack (YORP) spin-up, coupled with low levels of cohesion, can cause the disaggregation of rubble pile asteroids.
\citet{scheeres2017} suggests that this process might lead to ``fundamental constituents'' (monolithic rocks), reaching a size limit in their distribution that depends on many parameters of the body (density, strength, etc.).
According to this model, S-complex bodies tend to fragment more into their fundamental constituents than C-complex objects,
possibly due to their drier nature. Hence, the lack of carbonaceous material that we see in the small NEO population could be due not only to an observational bias, but could also have a physical nature related to their higher mechanical resistance to rotational fission.
One of the unknowns in the model in \citet{scheeres2017} is the size of these fundamental constituents for the different taxonomic types. Our results pointed out that the limit size for C-complex asteroids should reside within our D < 300 size bin, where we start to
observe a differentiation in terms of relative number of objects within the S and C populations. It is worth noting again
that these conclusions are  based on a small data set and therefore  will be investigated in detail during the second year of the project.

In our survey we classified only one A-type asteroid, which is in the small diameter range. This could be related to what is known as the “missing olivine problem”: A-type olivine-dominated asteroids are rare in both asteroidal surveys and meteorite collections (Sanchez et al. \citeyear{sanchez2014}). Our results might indicate that most of the olivine material in the solar system could have been battered to bits, making them elusive to be found. However, due to the limited number of A-type objects in our survey, more observations of A-type objects are necessary to confirm this preliminary result.

\section{Conclusions and future perspectives}

During the first year of the NEOShield-2 project we obtained B-V-R-I color indexes for 67 uncharacterized NEOs. We derived a  taxonomic classification for all of them. The overall majority belongs to the S-complex (41 objects), while the C- and X-complex account for 11 and 7 objects, respectively. Other taxonomic types (V-, D- and A-type) account for the remaining 8 objects.

The B-R versus V-I color diagram shows that the defined taxonomic groups occupy different regions, and confirms the goodness of our taxonomy obtained with the M4AST tool; our sample is also in agreement with the typical ranges found for C, S, and X-complex objects using the sample spectra from the Bus-DeMeo taxonomy. 
The analysis of the orbital parameters of our targets (a, e, i, q, Q, MOID) confirms that carbonaceous material is more related to the middle/outer main belt, while silicate NEOs show a connection with inner main belt. 

The analysis of the size distribution among taxonomical classes has pointed out that there seems to be a lack of carbonaceous material going to smaller sizes, and that the only olivine-dominated asteroid lies in this size range. While it is possible that a physical phenomenon is responsible for the lack of smaller carbonaceous NEOs, observational selection effects can still play a significant role. More observations of the small NEO population are needed at different  wavelength ranges in order to settle the question of what is causing these selection effects.  
For these reasons in the second year of the project we will focus more our observational campaign on smaller and fainter NEO targets (H > 20). Other than revealing whether the lack of carbonaceous material at small size ranges is due to a physical limit or whether it is bias-driven, and  revealing important information on the formation and evolution of the solar system (i.e., the missing olivine problem), the physical characterization of small NEOs (D < 300 m) is crucial, due to the higher likelihood of their potential impact on our planet. The characterization of  their mitigation-relevant quantities (e.g., surface composition, albedo, size) will  therefore be fundamental  to designing a successful mitigation strategy.

The current knowledge of mitigation missions is still at the very beginning, and this is particularly true for porous carbonaceous asteroids. Only few laboratory experiments have assessed the role of porosity in the mitigation of threatening asteroids. Moreover, while our knowledge of mineralogy of asteroids has greatly improved over the last ten years, we know very little about their internal structure.  In the next years, future space missions planned toward carbonaceous NEOs will determine more about their structure, while new laboratory experiments should assess  the role of porosity in the mitigation of potential impactors  in greater detail.
New  observations of the small NEO population, performed in synergy with future space missions and laboratory experiments, will not only  improve the design of future mitigation missions, but will also allow us to reach a new level of comprehension of the current unsolved issues of the small bodies population.

\begin{acknowledgements}
This research has been funded with support from the European Commission (grant agreement no. 640351 H2020- PROTEC-2014 - Access technologies and characterization for Near Earth Objects (NEOs). 
SI also acknowledges financial support from ASI (contract No. 2013-046-R.0: ``OSIRIS-REx Partecipazione Scientifica alla missione per  la fase B2/C/D'').
DP has received funding from the European Union's Horizon 2020 research and innovation program under the Marie Sklodowska-Curie actions (grant agreement n. 664931).
We would also like to thank the anonymous referee for the insightful comments that helped  improve the paper.
\end{acknowledgements}

%
%

\bibliographystyle{aa} 
\bibliography{reference2} 


\newpage 
\longtab{
\begin{longtable}{l c c c r} 
\caption{Observational circumstances for the  sample of 67 NEOs observed.} \\
\label{observations}      
Object                  & Date & $\Delta$ &   r   & $\alpha$\\
                       &         &   (au)   & (au)  &     \\  
\hline
\endfirsthead
\caption{continued.}\\
\hline\hline
Object                  & Date & $\Delta$ &   r   & $\alpha$  \\
                       &         &   (au)   & (au)  &     \\  
\hline
\endhead
\hline
\endfoot
4596    1981 QB &       12/02/2016      &       1.813   &       2.573   &       16.8    \\
5370    Taranis &       02/05/2016      &       2.301   &       3.250   &       7.2     \\
6050    Miwablock*      &       12/02/2016      &       2.458   &       3.129   &       15.0    \\
        &       02/05/2016      &       2.340   &       3.165   &       12.2    \\
138852  2000 WN10       &       10/12/2015      &       0.289   &       1.197   &       37.9    \\
142563  2002 TR69       &       10/12/2015      &       1.131   &       1.804   &       29.1    \\
154275  2002 SR41       &       01/07/2014      &       0.224   &       1.182   &       38.8    \\
155110  2005 TB &       10/12/2015      &       0.782   &       1.526   &       34.4    \\
159857  2004 LJ1        &       01/07/2014      &       1.256   &       2.080   &       21.2    \\
162273  1999 VL12       &       10/12/2015      &       1.008   &       1.837   &       22.6    \\
174806  2003 XL &       10/12/2015      &       1.098   &       1.965   &       18.3    \\
194126  2001 SG276      &       10/12/2015      &       0.874   &       1.756   &       20.3    \\
242708  2005 UK1        &       01/07/2014      &       0.535   &       1.462   &       27.3    \\
243566  1995 SA &       01/07/2014      &       0.532   &       1.239   &       53.5    \\
250706  2005 RR6        &       06/04/2016      &       0.689   &       1.512   &       32.4    \\
267136  2000 EF104      &       06/04/2016      &       0.479   &       1.448   &       17.4    \\
333578  2006 KM103      &       02/07/2016      &       0.360   &       1.223   &       47.8    \\
334412  2002 EZ2*       &       06/04/2016      &       0.318   &       1.306   &       14.3    \\
        &       02/05/2016      &       0.403   &       1.302   &       36.9    \\
334673         2003 AL18    &   02/12/2015      &       0.791   &       1.722   &       15.6    \\
356285  2010 DE &       06/04/2016      &       1.012   &       1.989   &       8.8     \\
411201  2010 LJ14       &       02/05/2016      &       0.614   &       1.433   &       36.5    \\
418416  2008 LV16       &       01/07/2014      &       0.231   &       1.206   &       31.7    \\
430804  2005 AD13       &       12/02/2016      &       1.031   &       1.996   &       8.4     \\
436771  2012 JG11*      &       03/06/2016      &       0.633   &       1.373   &       43.7    \\
        &       02/07/2016      &       0.608   &       1.370   &       43.3    \\
441987  2010 NY65       &       01/07/2014      &       0.080   &       1.033   &       76.2    \\
442243  2011 MD11       &       10/12/2015      &       0.808   &       1.559   &       33.0    \\
443880  2001 UZ16       &       10/12/2015      &       0.446   &       1.329   &       33.0    \\
445974  2013 BJ18       &       13/10/2015      &       0.690   &       1.683   &       5.3     \\
447221  2005 UO5        &       02/05/2016      &       0.272   &       1.114   &       60.6    \\
451370  2011 AK5        &       13/10/2015      &       0.192   &       1.165   &       27.2    \\
452397         2002 PD130    &          02/12/2015      &       0.492   &       1.474   &       5.1     \\
453729  2011 BO24*      &       12/02/2016      &       0.622   &       1.453   &       32.6    \\
        &       06/04/2016      &       0.490   &       1.231   &       51.3    \\
455199  2000 YK4        &       10/12/2015      &       0.479   &       1.380   &       28.3    \\
457663  2009 DN1        &       02/07/2016      &       0.284   &       1.156   &       54.2    \\
458375  2010 WY8*       &       12/02/2016      &       0.249   &       1.209   &       24.3    \\
        &       06/04/2016      &       0.274   &       1.206   &       37.2    \\
458723  2011 KQ12       &       12/02/2016      &       0.414   &       1.228   &       46.1    \\
459046  2012 AS10    &          02/12/2015      &       1.354   &       2.328   &       4.7     \\
463282  2012 HR15       &       06/04/2016      &       1.158   &       1.816   &       30.2    \\
466508  2014 GY48       &       06/04/2016      &       0.582   &       1.198   &       56.4    \\
467527  2007 LA15       &       02/05/2016      &       0.697   &       1.680   &       12.1    \\
468005  2012 XD112      &       10/12/2015      &       0.274   &       1.237   &       20.4    \\
468452  2003 SD170      &       02/07/2016      &       0.338   &       1.195   &       50.9    \\
468468  2004 KH17       &       03/06/2016      &       0.114   &       1.055   &       66.2    \\
468540  2006 MD12       &       02/07/2016      &       0.452   &       1.290   &       44.2    \\
468741  2010 VM1        &       02/07/2016      &       0.234   &       1.177   &       42.4    \\
469722  2005 LP40*      &       03/06/2016      &       0.530   &       1.288   &       48.2    \\
        &       02/07/2016      &       0.338   &       1.069   &       72.1    \\
474238  2001 RU17       &       02/09/2016      &       0.164   &       1.167   &       15.0    \\
482796  2013 QJ10       &       13/10/2015      &       0.531   &       1.368   &       37.2    \\
503277  2015 RT83       &       10/12/2015      &       0.301   &       1.159   &       48.2    \\
        1994 CJ1        &       01/07/2014      &       0.092   &       1.063   &       57.8    \\
        1999 VR6        &       03/06/2016      &       0.317   &       1.292   &       25.4    \\
        2001 UG18       &       30/07/2016      &       0.302   &       1.256   &       32.7    \\
        2009 DL46*      &       06/04/2016      &       0.159   &       1.048   &       68.5    \\
                &       02/05/2016      &       0.080   &       1.012   &       85.0    \\
        2009 VY25       &       02/09/2016      &       0.293   &       1.241   &       33.1    \\
        2012 BF86       &       12/02/2016      &       0.106   &       1.067   &       39.7    \\
        2012 CK2        &       12/02/2016      &       0.657   &       1.549   &       24.1    \\
        2012 JR17       &       02/05/2016      &       1.095   &       2.077   &       8.7     \\
        2014 RC11       &       02/09/2016      &       0.370   &       1.333   &       24.5    \\
        2015 OL35       &       12/02/2016      &       0.692   &       1.549   &       27.4    \\
        2015 OS35    &          02/12/2015      &       0.675   &       1.649   &       7.8     \\
        2015 TX143      &       02/05/2016      &       0.293   &       1.086   &       66.8    \\
        2015 YN1        &       12/02/2016      &       0.246   &       1.172   &       37.0    \\
        2016 AF165      &       12/02/2016      &       0.343   &       1.311   &       16.7    \\
        2016 EM28       &       02/05/2016      &       0.382   &       1.289   &       36.5    \\
        2016 GN221      &       02/07/2016      &       0.258   &       1.226   &       32.2    \\
        2016 LY47       &       02/07/2016      &       0.420   &       1.280   &       43.5    \\
        2016 LZ10       &       30/07/2016      &       0.197   &       1.156   &       40.7    \\
        2016 NV &       30/07/2016      &       0.415   &       1.268   &       44.6    \\
\end{longtable}
NOTE: $\Delta$ and $r$  are the topocentric and the heliocentric distance, respectively.
$\alpha$ is the solar phase angle. Objects with an asterisk were observed twice, as explained in the text.
}

\longtab{
\begin{longtable}{|l |c| c| l c c| c |c |c|} \\
\caption{Absolute (H) and visual (V) magnitude, color indexes, and the obtained taxonomy for our sample of 67 NEOs. Known albedos are taken form the EARN database, while diameters are computed using the relation found by \citet{bowell1989}.  } \\
\label{colors}      
Object  &               H       &               V               &       B-V     &       V-R     &       V-I     &       Taxonomy        &       Albedo  &       D (km)\\
\hline
\endfirsthead
\caption{continued.}\\
\hline\hline
Object  &               H       &               V               &       B-V     &       V-R     &       V-I     &       Taxonomy        &       Albedo  &       D (km)    \\
\hline
\endhead
\hline
\endfoot 
4596    &       16.3    &       $       21.13 \pm 0.06  $       &       0.95    &       0.47    &       0.96    &       S       &               &               \\
5370    &       15.1    &       $       20.10 \pm 0.03  $       &       0.64    &       0.36    &       0.82    &       C       &       0.040   &       6.060   \\
6050    &       14.9    &       $       20.44 \pm 0.05  $       &       0.76    &       0.37    &       0.73    &       X       &       0.186   &       3.227   \\
        &               &       $       20.11 \pm 0.04  $       &       0.73    &       0.42    &       0.74    &               &               &               \\
138852  &       20.2    &       $       19.57 \pm 0.04  $       &       0.83    &       0.48    &       0.69    &       S       &               &               \\
142563  &       17.1    &       $       19.88 \pm 0.03  $       &       0.92    &       0.43    &       0.61    &       S       &       0.378   &       0.822   \\
154275* &       20.1    &       $       18.67 \pm 0.03  $       &       0.8     &       0.41    &       0.71    &       C       &               &               \\
155110  &       17.5    &       $       19.10 \pm 0.03  $       &       0.77    &       0.31    &       0.74    &       S       &               &               \\
159857* &       15.4    &       $       18.63 \pm 0.04  $       &       0.78    &       0.58    &       0.9     &       S       &       0.130   &       3.066   \\
162273  &       17.2    &       $       19.51 \pm 0.05  $       &       0.86    &       0.55    &       0.84    &       V       &               &               \\
174806  &       17.2    &       $       19.96 \pm 0.08  $       &       0.76    &       0.45    &       0.95    &       D       &               &               \\
194126  &       17.7    &       $       20.03 \pm 0.05  $       &       0.75    &       0.55    &       0.87    &       S       &               &               \\
242708* &       18.1    &       $       18.96 \pm 0.06  $       &       0.74    &       0.5     &       0.72    &       S       &               &               \\
243566* &       17.4    &       $       18.21 \pm 0.06  $       &       0.77    &       0.53    &       0.79    &       S       &       0.091   &       1.459   \\
250706* &       18.5    &       $       19.46 \pm 0.06  $       &       0.83    &       0.48    &       0.82    &       S       &               &               \\
267136  &       18.9    &       $       18.98 \pm 0.04  $       &       0.83    &       0.46    &       0.72    &       S       &               &               \\
333578* &       20.2    &       $       19.74 \pm 0.03  $       &       0.82    &       0.47    &       0.65    &       V       &               &               \\
334412  &       20.2    &       $       18.95 \pm 0.05  $       &       0.69    &       0.45    &       0.76    &       S       &       0.400   &       0.192   \\
                &               &       $       20.37 \pm 0.05  $       &       0.76    &       0.42    &       0.81    &               &                &               \\      
334673  &       17.9    &       $       19.89 \pm 0.08  $       &       0.79    &       0.48    &       0.72    &       S       &       0.295   &       0.643   \\
356285  &       17.3    &       $       19.69 \pm 0.04  $       &       0.77    &       0.58    &       0.8     &       S       &               &               \\
411201  &       17.8    &       $       19.54 \pm 0.02  $       &       0.91    &       0.56    &       0.74    &       V       &       0.189   &       0.842   \\
418416* &       20.3    &       $       18.51 \pm 0.04  $       &       0.94    &       0.51    &       0.72    &       S       &               &               \\
430804* &       17.9    &       $       20.26 \pm 0.04  $       &       0.51    &       0.29    &       0.69    &       C       &       0.130   &       0.970   \\
436771  &       19.0    &       $       20.55 \pm 0.03  $       &       0.82    &       0.49    &       0.87    &       S       &               &               \\
                &               &       $       20.39 \pm 0.04  $       &       0.86    &       0.45    &       0.82    &               &               &               \\
441987* &       21.5    &       $       18.63 \pm 0.02  $       &       0.77    &       0.56    &       0.91    &       S       &       0.071   &       0.250   \\
442243  &       18.1    &       $       19.98 \pm 0.05  $       &       0.85    &       0.52    &       0.81    &       S       &               &               \\
443880* &       19.4    &       $       20.20 \pm 0.06  $       &       0.57    &       0.44    &       0.61    &       C       &               &               \\
445974* &       20.3    &       $       21.72 \pm 0.16  $       &       0.92    &       0.56    &       0.77    &       S       &               &               \\
447221  &       20.7    &       $       20.31 \pm 0.03  $       &       0.79    &       0.54    &       0.9     &       S       &               &               \\
451370* &       21.5    &       $       19.42 \pm 0.07  $       &       0.8     &       0.35    &       0.73    &       C       &               &               \\
452397  &       20.4    &       $       20.31 \pm 0.08  $       &       0.85    &       0.54    &       0.77    &       S       &               &               \\
453729* &       18.8    &       $       20.21 \pm 0.03  $       &       0.88    &       0.39    &       0.83    &       S       &               &               \\
                &               &       $       19.84 \pm 0.04  $       &       0.81    &       0.46    &       0.74    &                &               &               \\
455199  &       19.8    &       $       20.45 \pm 0.08  $       &       0.81    &       0.41    &       0.89    &       D       &               &               \\
457663  &       20.3    &       $       19.58 \pm 0.04  $       &       0.8     &       0.38    &       0.67    &       X       &               &               \\
458375  &       21.3    &       $       19.94 \pm 0.05  $       &       0.77    &       0.51    &       0.66    &       S       &               &               \\
                &               &       $       20.12 \pm 0.05  $       &       0.74    &       0.55    &       0.69    &                &               &               \\      
458723* &       19.4    &       $       19.49 \pm 0.03  $       &       1.24    &       0.5     &       0.68    &       V       &               &               \\
459046  &       17.1    &       $       20.43 \pm 0.08  $       &       0.94    &       0.56    &       0.91    &       S       &               &               \\
463282  &       17.1    &       $       19.82 \pm 0.05  $       &       0.72    &       0.49    &       0.72    &       X       &               &               \\
466508* &       18.6    &       $       20.08 \pm 0.05  $       &       0.73    &       0.57    &       0.79    &       S       &               &               \\
467527  &       19.5    &       $       20.59 \pm 0.03  $       &       0.81    &       0.38    &       0.79    &       X       &               &               \\
468005* &       21.2    &       $       19.78 \pm 0.04  $       &       0.99    &       0.5     &       0.69    &       S       &               &               \\
468452  &       19.1    &       $       18.86 \pm 0.03  $       &       0.74    &       0.33    &       0.63    &       C       &               &               \\
468468* &       21.9    &       $       19.28 \pm 0.01  $       &       0.78    &       0.54    &       0.65    &       S       &       0.072   &       0.206   \\
468540  &       19.4    &       $       20.12 \pm 0.02  $       &       0.76    &       0.48    &       0.71    &       S       &       0.430   &       0.267   \\
468741  &       20.1    &       $       18.60 \pm 0.02  $       &       0.76    &       0.42    &       0.69    &       S       &               &               \\
469722  &       19.5    &       $       20.45 \pm 0.05  $       &       0.96    &       0.39    &       0.80    &       S       &               &               \\
                &               &       $       20.18 \pm 0.03  $       &       0.91    &       0.45    &       0.85    &               &               &               \\
474238  &       21.5    &       $       18.80 \pm 0.02  $       &       0.75    &       0.44    &       0.8     &       X       &               &               \\
482796* &       19.5    &       $       20.19 \pm 0.04  $       &       0.95    &       0.52    &       0.73    &       S       &               &               \\
503277  &       19.8    &       $       19.30 \pm 0.10  $       &       0.94    &       0.39    &       0.65    &       S       &               &               \\
1994 CJ1*       &       21.4    &       $       18.55 \pm 0.04  $       &       0.83    &       0.72    &       1.02    &       A       &               &               \\
1999 VR6*       &       20.8    &       $       20.15 \pm 0.09  $       &       0.8     &       0.5     &       0.74    &       S       &               &               \\
2001 UG18       &       20.7    &       $       19.88 \pm 0.05  $       &       0.91    &       0.41    &       0.59    &       S       &               &               \\
2009 DL46*      &       22.0    &       $       20.07 \pm 0.03  $       &       0.65    &       0.46    &       0.87    &       D       &               &               \\
                &               &       $       19.26 \pm 0.03  $       &       0.64    &       0.49    &       0.9     &                &               &               \\
2009 VY25       &       19.6    &       $       18.82 \pm 0.03  $       &       0.61    &       0.37    &       0.69    &       C       &               &               \\
2012 BF86       &       22.6    &       $       19.87 \pm 0.04  $       &       0.93    &       0.43    &       0.59    &       S       &               &               \\
2012 CK2        &       18.9    &       $       19.61 \pm 0.03  $       &       0.7     &       0.38    &       0.59    &       C       &               &               \\
2012 JR17       &       17.7    &       $       19.72 \pm 0.02  $       &       0.81    &       0.41    &       0.83    &       X       &               &               \\
2014 RC11       &       20.7    &       $       20.30 \pm 0.04  $       &       0.86    &       0.52    &       0.88    &       S       &               &               \\
2015 OL35*      &       18.0    &       $       19.40 \pm 0.04  $       &       0.77    &       0.52    &       0.75    &       S       &               &               \\
2015 OS35       &       19.0    &       $       19.47 \pm 0.06  $       &       0.82    &       0.28    &       0.61    &       C       &               &               \\
2015 TX143      &       19.8    &       $       19.58 \pm 0.02  $       &       0.77    &       0.52    &       0.72    &       S       &               &               \\
2015 YN1        &       21.0    &       $       20.01 \pm 0.05  $       &       0.78    &       0.59    &       0.85    &       S       &               &               \\
2016 AF165      &       20.2    &       $       19.30 \pm 0.03  $       &       0.67    &       0.33    &       0.67    &       C       &               &               \\
2016 EM28       &       20.7    &       $       21.00 \pm 0.02  $       &       0.72    &       0.37    &       0.78    &       X       &               &               \\
2016 GN221      &       20.5    &       $       19.23 \pm 0.06  $       &       0.88    &       0.4     &       0.76    &       X       &               &               \\
2016 LY47       &       19.8    &       $       19.84 \pm 0.04  $       &       0.81    &       0.5     &       0.72    &       S       &               &               \\
2016 LZ10       &       20.0    &       $       17.91 \pm 0.04  $       &       0.8     &       0.4     &       0.65    &       C       &               &               \\
2016 NV*        &       19.8    &       $       20.01 \pm 0.03  $       &       0.93    &       0.5     &       0.79    &       S       &               &               \\
\end{longtable}
NOTE: Objects with an asterisk are potentially hazardous asteroids (PHAs), as defined in the text.
}

  \begin{table}
\caption{B-R and V-I color ranges found in our sample and using the DeMeo sample spectra. 
Our colors are computed from the median value and using the 1-$\sigma$ deviation as the error. DeMeo colors are computed integrating the flux of the sample spectra from DeMeo et al. (2009) over the transmission range of each filter. Errors in this case take into account the dispersion of these sample spectra.
}             
\label{colorranges}      
\centering                          
\begin{tabular}{|l |c |c| c|c|}        
\hline             
& $ (B-R)_ {Ieva17}$& $(B-R)_ {DeMeo} $  &$(V-I)_ {Ieva17}$ &$(V-I)_ {DeMeo}$\\  
\hline  
C&$1.07 \pm 0.12$ & $1.09^{+0.11}_ {-0.09}$& $0.67 \pm 0.07$&$0.69^{+0.04}_ {-0.10}$\\ 
\hline
D& $1.21 \pm 0.06$ & $1.18^{+0.07}_ {-0.06}$& $0.89 \pm 0.03$&$0.89^{+0.06}_ {-0.07}$\\ 
\hline
S& $1.32 \pm 0.07$ & $1.29^{+0.14}_ {-0.11}$& $0.77 \pm 0.09$&$0.82^{+0.10}_ {-0.14}$\\ 
\hline
V&$1.44 \pm 0.19$ & $1.34^{+0.09}_ {-0.09}$& $0.71 \pm 0.08$&$0.75^{+0.08}_ {-0.08}$\\ 
\hline 
X& $1.19 \pm 0.06$ & $1.13^{+0.12}_ {-0.07}$& $0.78 \pm 0.05$&$0.78^{+0.06}_ {-0.04}$\\ 
\hline
\end{tabular}
\end{table}

\begin{table}
\caption{Median orbital parameters (semimajor axis, eccentricity, inclination, aphelion, perihelion, and minimum orbital intersection distance) for the two major groupings considered in our analysis}             
\label{orbital}      
\centering                          
\begin{tabular}{|l |c|c|}        
\hline                
&C (11)&S (41)\\
\hline                        
a (au)& $2.21 \pm 0.72$ & $1.61 \pm 0.59$               \\
e& $0.49 \pm 0.13$      & $0.43 \pm 0.17$               \\
i (deg)&  $11.58 \pm 7.15$ & $15.73 \pm 9.53$           \\
q (au)&  $1.14 \pm 0.28$        & $0.93 \pm 0.26$               \\
Q (au)& $3.42 \pm 1.26$ & $2.28 \pm 1.09
$               \\
MOID (au)& $0.15 \pm 0.15$      & $0.11 \pm 0.11$               \\
\hline                                   
\end{tabular}
\end{table}

  \begin{figure}
 \centering
   \includegraphics[angle=0,width=9cm]{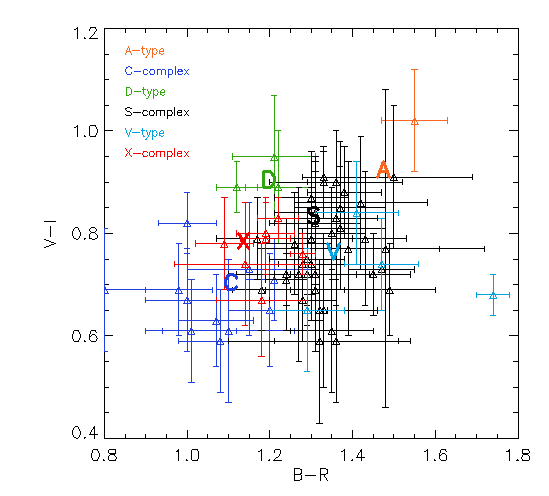}
      \caption{B-R vs. the V-I color indexes for  the sample considered in our analysis (see Table 2). Also shown (in bold) are the average values obtained from the DeMeo et al. (2009) sample spectra with the method described above (see also Table 3).
}
        \label{colors}
   \end{figure}

     \begin{figure}
  \centering
   \includegraphics[angle=0,width=18cm]{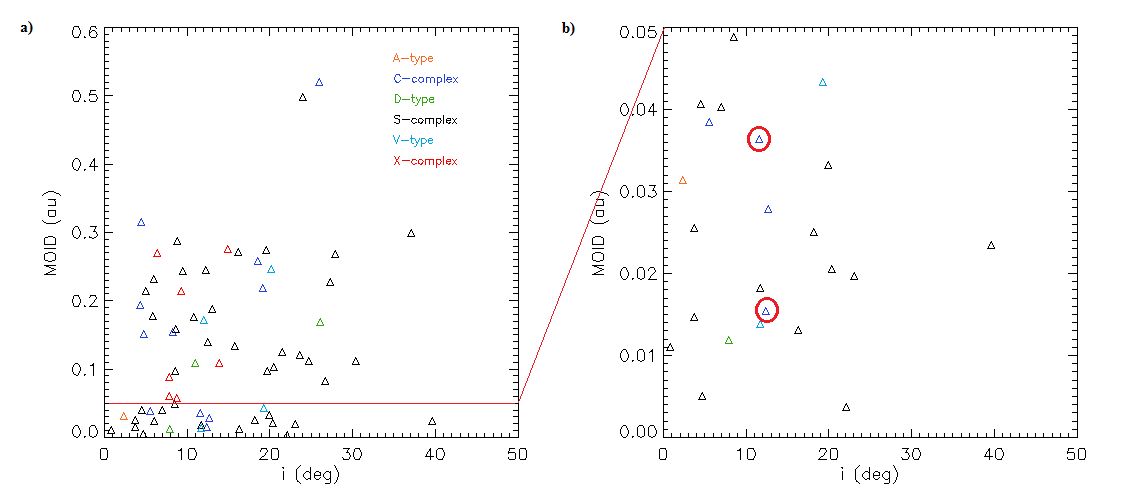}
      \caption{a) Inclination vs. MOID reported for the whole sample of NEOs and b) zoom on MOID < 0.05, which defines the PHA population (23 bodies in our sample). In particular, two carbonaceous targets (circled in red) show very low MOID and very high inclination. These objects are on a challenging orbit for mitigation purposes, and require a detailed analysis.
 }
        \label{moidinc}
   \end{figure}

    \begin{figure}
   \centering
   \includegraphics[angle=0,width=18cm]{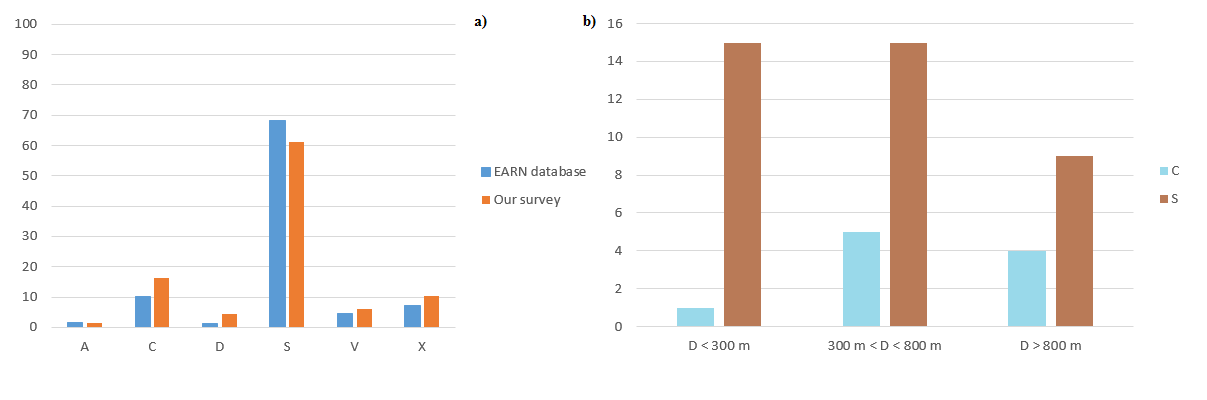}
      \caption{a) Percentage distribution of the taxonomic complexes and classes in our survey compared to those reported in the EARN database (www.earn.dlr.de/nea/table1\textunderscore{new}.html) and b) the distribution of C- and S-complex objects classified in our survey in three different size ranges: D < 300 m, 300 m < D < 800 m, and D > 800 m.
}
        \label{earn}
   \end{figure}

\end{document}